\documentclass{article}
\usepackage{spconf,amsmath,graphicx,hyperref}
\usepackage{booktabs}
\usepackage{multirow}
\usepackage{array}
\usepackage{tcolorbox}

\usepackage{enumitem}

\usepackage{threeparttable}

\usepackage{makecell}

\usepackage{siunitx}
\usepackage{tabularx}
\usepackage[skip=3pt]{caption} 
\usepackage{arydshln}

\title{ABC-Eval: Benchmarking Large Language Models\\
on Symbolic Music Understanding and Instruction Following}
\name{
Jiahao Zhao$^{1}$\quad
Yunjia Li$^{2}$\quad
Wei Li$^{2}$\quad
Kazuyoshi Yoshii$^{3}$
\thanks{This work was supported 
by JST FOREST No. JPMJFR2270 
and JSPS KAKENHI Nos. 24H00742, 24H00748, 25K22841, and 25H01142.}
}
\address{
  $^{1}$Graduate School of Informatics, Kyoto University, Japan\\
  $^{2}$College of Computer Science and Artificial Intelligence, Fudan University, China\\
  $^{3}$Graduate School of Engineering, Kyoto University, Japan\\}
\begin{document}
%\ninept
\fussy
\maketitle
\begin{abstract}
As large language models (LLMs) continue to develop, the feasibility and significance of text-based symbolic music tasks have become increasingly prominent. While symbolic music has been widely used in generation tasks, LLM capabilities in understanding and reasoning about symbolic music remain largely underexplored.
To address this gap, we propose ABC-Eval, the first open-source benchmark dedicated to the understanding and instruction-following capabilities in text-based ABC notation scores. It comprises 1,086 test samples spanning 10 sub-tasks, covering scenarios from basic musical syntax comprehension to complex sequence-level reasoning. Such a diverse scope poses substantial challenges to models’ ability to handle symbolic music tasks.
We evaluated seven state-of-the-art LLMs on ABC-Eval, and the results reveal notable limitations in existing models’ symbolic music processing capabilities. Furthermore, the consistent performance of individual baselines across different sub-tasks supports the reliability of our benchmark.
\end{abstract}

\vspace{-0.7em}
\section{Introduction}
\vspace{-0.7em}
Symbolic music, often regarded as the \textit{language of music}~\cite{symbolicmusic}, encodes musical information into discrete, human-readable or machine-interpretable symbols. This form of representation enables diverse applications from digital preservation to computational analysis~\cite{mirsurvey}. The discrete nature of symbolic music allows it to be tokenized and processed by language models. Specialized music language models (MLMs) such as MusicBERT~\cite{musicbert} and Pianobart~\cite{bart} have thus demonstrated strong performance in various downstream tasks. 
\begin{figure}[!t]
    \centering
    \includegraphics[width=.9\columnwidth,trim=5mm 200mm 10mm 7mm, clip]{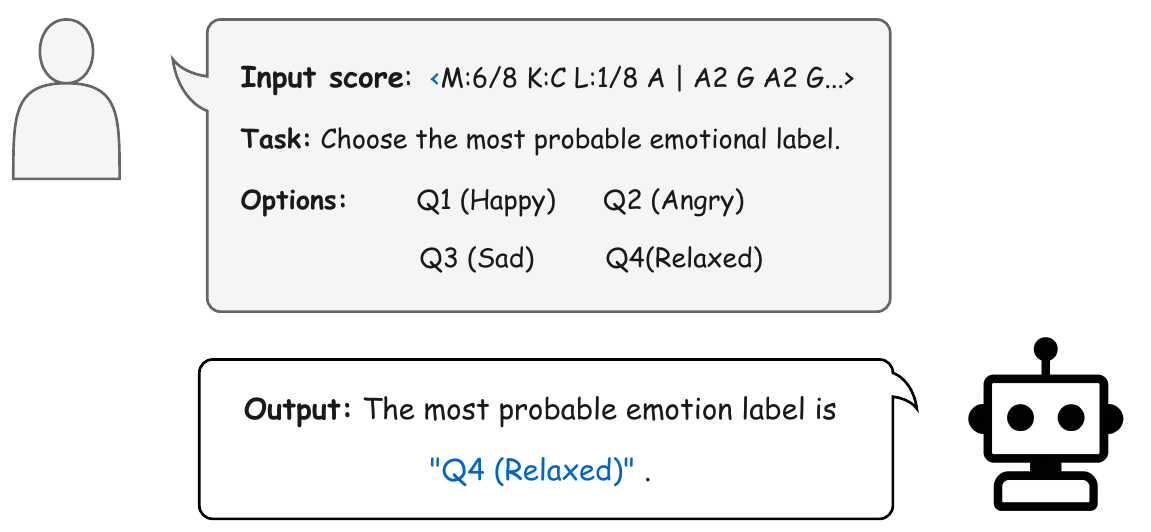}
    \caption{A conceptual example of input instructions and corresponding LLM output in ABC-Eval benchmark.}
    \label{fig1}
    \vspace{-1em}
\end{figure}

The success of large language models~\cite{llmoverview} (LLMs) in natural language processing has naturally enabled their extension to music-related applications. Since the pre-training corpora of LLMs typically contain large amounts of music-related text and raw symbolic music data, LLMs have exhibited capabilities in symbolic music tasks~\cite{llmmusic}. Recent studies have shown that LLMs can attain competitive performance in music question-answering (Music QA)~\cite{chatmusician,musicqa2} and conditional music generation~\cite{symbolicgenerationsurvey}. However, the symbolic music understanding and instruction-following capabilities of LLMs remain unexplored. A comprehensive and challenging benchmark is urgently needed to systematically evaluate and thereby improve LLMs' symbolic music processing capabilities.
\vspace{-0.02em}

To address this issue, we introduce ABC-Eval, the first benchmark for symbolic music understanding and instruction-following that is specifically designed for text-based LLMs. 
The benchmark uses ABC notation as its input format and comprises 10 sub-tasks with 1,086 test samples in total. A conceptual example of ABC-Eval test samples is shown in Fig~\ref{fig1}.
We release this benchmark for research and academic use to facilitate progress in this emerging area.\footnote{\url{https://anonymous.4open.science/r/ABC-Eval-B622}} 

Another contribution is to design a diverse and comprehensive set of evaluation tasks that span multiple levels of musical understanding and instruction-following. Our benchmark includes both segment-level and sequence-level tasks, ranging from fundamental musical syntax parsing to high-level musical semantic reasoning. The evaluation protocol encompasses standard multiple-choice questions as well as more challenging protocols with structured output that require precise instruction-following capability. 

We conduct extensive experiments on ABC-Eval using seven state-of-the-art LLMs, including both lightweight and ultra-large models. We demonstrate that the benchmark poses significant challenges even for the most advanced models, highlighting a critical gap between LLMs' theoretical music knowledge and their practical symbolic music reasoning capabilities.

\begin{comment}
\begin{itemize}
\setlength{\itemsep}{0pt}
\setlength{\parskip}{0.5pt}
\item We introduce ABC-Eval, the first symbolic music understanding and instruction-following benchmark specifically designed for text-based LLMs. The benchmark uses ABC notation as its input format and comprises 10 sub-tasks with 1086 test samples in total. We release this benchmark for research and academic use to facilitate progress in this emerging area.
\item We design a diverse and comprehensive set of evaluation tasks that span multiple levels of musical understanding and instruction-following. Our benchmark includes both segment-level and sequence-level tasks, ranging from fundamental syntax parsing to high-level musical semantic reasoning. The evaluation protocol encompasses standard multiple-choice questions as well as more challenging protocol with structured output that require precise instruction-following capability.
\item We conduct extensive experiments on ABC-Eval using 8 state-of-the-art LLM baselines, including both open-source LLMs and commercial API models. Our results demonstrate that the benchmark poses significant challenges even for the most advanced models, highlighting a critical gap between LLMs' theoretical music knowledge and their practical symbolic music reasoning capabilities.
\end{itemize}
\end{comment}

\vspace{-2mm}
\section{Related Work}
\vspace{-2mm}

%We reviews related work on 
%\subsection{Music understanding benchmarks}

The evaluation of music understanding capabilities have primarily relied on task-specific datasets targeting individual musical dimensions. MoodyLyrics~\cite{moodylyrics} and PMEmo~\cite{pmemo} datasets,
 for example, were developed to assess emotion recognition capabilities, while GTZAN~\cite{gtzan} and MagnaTagATune~\cite{magnatagatune} datasets focused on genre classification tasks. These benchmarks, though valuable for their purposes, offered limited insight about comprehensive music understanding abilities.

Recent benchmarks have shifted toward comprehensive evaluation frameworks that assess multiple facets through diverse tasks and protocols. Li et al.~\cite{ziqi} presented ZIQI-Eval, based on a manually-curated music question-answering (music QA) dataset that evaluates LLMs across various musical domains. Their findings revealed that LLMs can surpass PhD-level human performance in areas such as music history and theory, highlighting the significant potential of language models in music-related tasks.
Building upon the work of ZIQI-Eval, several multimodal music understanding benchmarks have been proposed to evaluate audio-language models. Weck et al.~\cite{mucho} introduced MuChoMusic, a comprehensive multimodal music QA dataset. Similarly, Ma et al.~\cite{cmibench} developed CMI-Bench, a benchmark that evaluates audio-text LLMs across traditional Music Information Retrieval (MIR) tasks. Notably, CMI-Bench emphasizes instruction-following capabilities and includes challenging note-level tasks such as beat tracking and melody extraction, which pose significant difficulties for current LLMs. 

While these benchmarks have substantially advanced our understanding of multimodal music comprehension, they primarily address scenarios involving audio inputs. Despite the advantages of text-based symbolic music in generation tasks, particularly in its high controllability and expressiveness~\cite{notagen}, the understanding capabilities of LLMs for text-encoded symbolic music remain largely unexplored. Mundada et al.~\cite{wildscore} recently made initial progress in this direction with WildScore, which combines textual descriptions with sheet music images to evaluate reasoning capabilities across musical dimensions including harmony, tonality, and texture. However, this approach still relies on visual representations rather than pure text-based symbolic notation. Our benchmark adopts ABC notation—a lightweight, human readable, and LLM-friendly representation—as the primary music input format, addressing the lack of text-based symbolic music benchmarks.
\renewcommand{\arraystretch}{1.1}

\begin{table}[t]
\centering

\caption{Statistical overview of the benchmark.}
\vspace{-1mm}
\label{tab:music_tasks}
\fontsize{9.5}{11}\selectfont % 强制9pt字号
\begin{tabularx}{\columnwidth}{p{2.35cm}!{\vrule width 0.8pt}X c}
\Xhline{0.8pt}
\textbf{Category} & \textbf{Sub-task} & \textbf{Test Num} \\
\Xhline{0.8pt}

% ----- Basic Syntax Understanding -----
\multirow{2}{*}{\makecell[tl]{\textbf{Basic Syntax}\\\textbf{Understanding}}}
  & Bar Count Estimation & 100 \\[-1pt]
  \cdashline{2-3}[0.5pt/2pt]
  & Metadata QA& 60 \\
\Xhline{0.6pt}

% ----- Segment-level Understanding -----
\multirow{3}{*}{\makecell[tl]{\textbf{Segment-level}\\\textbf{Understanding}}}
  & Next-Bar Prediction & 119 \\[-2pt]
  \cdashline{2-3}[0.5pt/2pt]
  & Bar Sequencing & 119 \\[-2pt]
  \cdashline{2-3}[0.5pt/2pt]
  & Error Detection (F1) & 220 \\
\Xhline{0.6pt}

% ----- Sequence-level Understanding -----
\multirow{5}{*}{\makecell[tl]{\textbf{Sequence-level}\\\textbf{Understanding}}}
  & Music Captioning & 60 \\[-2pt]
  \cdashline{2-3}[0.5pt/2pt]
  & Metadata Prediction & 60 \\[-2pt]
  \cdashline{2-3}[0.5pt/2pt]
  & Emotion Recognition & 120 \\[-2pt]
  \cdashline{2-3}[0.5pt/2pt]
  & Composer Recognition & 96 \\[-2pt]
  \cdashline{2-3}[0.5pt/2pt]
  & Genre Recognition & 132 \\
\Xhline{0.6pt}

\textbf{Total Num} &  & \textbf{1086} \\
\Xhline{0.8pt}
\end{tabularx}
\end{table}

\vspace{-2mm}
\section{The ABC-Eval Benchmark}
\vspace{-2mm}

This section presents the data sources, task categories, and corresponding sub-tasks of ABC-Eval. For each sub-task, we describe its task, evaluation protocol, and metrics employed. Note that all multiple-choice tasks use accuracy as the evaluation metric. The statistical overview of the benchmark is summarized in Table~\ref{tab:music_tasks}.

\vspace{-2mm}
\subsection{Data sources}
\vspace{-1mm}

For most tasks, we employ rule-based data construction using ABC scores from Nottingham~\cite{nottingham} Music dataset and Irish Massive ABC Notation (IrishMAN)~\cite{irishman} dataset, selected for their public availability and high annotation quality. We utilize the bar count field from the IrishMAN dataset and the title field from the Nottingham dataset as ground truth for bar count estimation and music captioning tasks, respectively. The Error Detection task requires construction based on high-quality samples without syntactic errors; therefore, 22 samples were manually selected by experts. For emotion, genre, and composer recognition tasks, we use EMOPIA~\cite{emopia}, ADL-piano MIDI~\cite{midibert}, and Pianist8~\cite{genre} as data sources, respectively. Since only MIDI files are available in these datasets, we employed a state-of-the-art MIDI-to-score model~\cite{miditoscore} to transcribe the MIDI files to MusicXML format, then used the EasyABC library to convert them to ABC notation.

All data in this benchmark is intended strictly for non-commercial research purposes. We strictly adhered to the licenses of the original data sources during open-source release and ensured ethical compliance and usability.

\vspace{-2mm}
\subsection{Basic syntax understanding tasks}
\vspace{-1mm}

Basic syntax understanding tasks are designed to assess models' fundamental understanding of ABC notation syntax and their basic instruction-following capabilities. These sub-tasks can serve as alignment datasets for model training.

\textbf{Metadata Question Answering}:
Models are required to select the correct option from four choices corresponding to the key signature, meter, or note length field of the input score. 

\textbf{Bar Count Estimation}:
Models are required to directly output the number of bars in the input score. Accuracy metrics are used to evaluate this task, where only outputs that exactly match the ground truth are considered correct during metric computation.

\subsection{Segment-level understanding tasks}

\textbf{Next-Bar Prediction}:
Given the first few bars of a score as input, models are required to select the most likely next bar segment from four options. The candidate options consist of the ground truth option and three randomly selected bars from the remaining portion of the current score.

\textbf{Bar Sequencing}:
This task is comparatively challenging. Given several randomly shuffled bars as input, models must output the correct sequential order using corresponding indices (e.g., 0312). We employ the Kendall tau coefficient to measure the ordinal association between the output and ground truth sequences. The resulting coefficient is mapped from the interval $[-1,1]$ to $[0,1]$ to facilitate averaging with metrics from other tasks. We multiply the coefficient by a completeness penalty factor to penalize outputs that are too short, while excessively long or rule-violating outputs are assigned the minimum value of 0.

\textbf{Error Detection}:
This task is highly challenging, primarily testing models' comprehension abilities and instruction-following capabilities in complex scenarios. Given a complete score containing multiple errors as input, models must identify all bar numbers containing errors, with the metadata section considered part of the first bar. Macro-F1 score is used to provide a balanced evaluation of both precision and recall.

We employ a rule-based approach to construct samples for this task. Since errors or ambiguities in original scores may affect task performance, 22 high-quality scores are manually selected by music experts as generation samples. We insert several random errors at random positions in the samples. We designed diverse error categories ranging from basic syntactic errors to music theory violations:
\begin{enumerate}[leftmargin=*, noitemsep]
\setlength{\itemsep}{0pt}
\setlength{\parskip}{0.2pt}
\item Invalid metadata, such as invalid key signatures and time signatures (e.g., M:6/7, K: Rmin)
\item Invalid content (insert non-existent pitch numbers and note values, e.g., 'Z8')
\item Invalid bar durations (e.g., M:4/4 but actual duration sum in a bar is 3/4)
\item Unreasonable melodic leaps (modifying a note to create a sudden interval jump $>$ 10 degrees)
\item Accidental errors outside the key signature (e.g., writing \texttt{\_C} in G major where flat C is not in the key signature)
\end{enumerate}
Note that error type 1 (invalid metadata) does not co-occur with types 3, 4, and 5 to avoid logical conflicts

\vspace{-1em}
\subsection{Sequence-level understanding tasks}
\textbf{Metadata Prediction}:
Models are required to select the right option for a specific metadata field of the input score from four choices. Unlike the previous Metadata QA task, we mask certain metadata fields in this task and require models to make judgments or predictions based on patterns within the score content.

\textbf{Music Captioning}:
Models are required to select the right title for the input score from four options. We use the title fields provided with samples in the Nottingham dataset as ground truth labels, and randomly select three additional titles from the complete title collection to form the candidate options.

\textbf{Emotion, Genre, and Composer Recognition}:
These tasks share the same structure and evaluation protocol. Models are required to select the right emotion/genre/composer label from four options based on the input score. The emotion label candidates correspond to the four quadrants of Russell's circumplex model used by the EMOPIA~\cite{emopia} dataset, while composer and genre label candidates are randomly selected from the label sets of the Pianist8~\cite{midibert} and ADL-piano MIDI~\cite{genre} datasets.
\begin{table}[t]
\centering
\caption{The prompt configurations used.}
\vspace{-2mm}
\begin{tcolorbox}[title=Prompt Template for Multiple-Choice Questions, coltitle=black, colback=white, colframe=black!30, fontupper=\small\ttfamily]
\textbf{Input}: \{\textit{input\_content}\}\\
\textbf{Task}: \{\textit{task\_instruction}\}\\
\textbf{Options}:\\
0. \{\textit{options[0]}\}
1. \{\textit{options[1]}\}\\
2. \{\textit{options[2]}\}
3. \{\textit{options[3]}\}\\\\
Please only output the index of the correct option (0, 1, 2, or 3), do not output any additional content.
\end{tcolorbox}
\vspace{0.1em}
\begin{tcolorbox}[title=Prompt Template for Structured-Output Questions, coltitle=black, colback=white, colframe=black!30, fontupper=\small\ttfamily]
\textbf{Input}: \{\textit{input\_content}\}\\
\textbf{Task}: \{\textit{task\_instruction}\}\\
\textbf{Template}: \{\textit{structured\_output\_template}\}\\\\
Please directly output the answer of the given task, without any explanation or additional content.
\end{tcolorbox}
\label{tab:prompt}
\vspace{-2em}
\end{table}
\renewcommand{\arraystretch}{1}
\begin{table*}[t]
\centering
\caption{Multi-task evaluation results across all baselines (best per row in \textbf{bold}). Bar Sequencing task and Error Detection task use Kendall's $\tau$ coefficient and Macro-F1 score as their metrics, respectively; other sub-tasks use classification accuracy. All metrics are reported in percentage points (pts, higher is better).  }
\label{tab:multi_task_eval}
\fontsize{9pt}{11pt}\selectfont
\begin{tabularx}{0.95\textwidth}{lXXXXXXX}
\toprule
\textbf{Tasks} & \textbf{DeepSeek-chat} & \textbf{DeepSeek-reasoner} & \textbf{Gemini-pro} & \textbf{Gemini-flash} & \textbf{GPT-5-mini} & \textbf{GPT-5-nano} & \textbf{GPT-5} \\
\midrule
Bar Count Estimation & 30.00 & 91.00 & 66.00 & 84.00 & \textbf{99.00} & 95.00 & 93.00 \\
Metadata QA          & \textbf{100} & 98.33 & 98.33 & 98.33 & \textbf{100} & 98.33 & \textbf{100} \\
\textbf{Basic Syntax Average}   & 65.00 & 94.67 & 82.17 & 91.17 & \textbf{99.50} & 96.67 & 96.5  \\
\midrule
Bar Sequencing (Kendall's $\tau$) & 46.41 & 53.75 & 59.01 & 44.85 & 56.33 & 48.80 & \textbf{59.66} \\
Next-Bar Prediction  & 32.77 & 42.86 & 43.70 & 26.89 & 43.70 & 38.66 & \textbf{50.42}  \\
Error Detection (Macro-F1) & 6.29 & 25.34 & \textbf{27.22} & 25.86 & 25.62 & 17.87 & 24.12  \\
\textbf{Segment-level Average}   & 28.49 & 40.65 & 43.31 & 32.53 & 41.88 & 35.11 & \textbf{44.73} \\
\midrule
Music Captioning     & 26.67 & 23.33 & 28.33 & 28.33 & \textbf{40.00} & 26.67 & 33.33 \\
Metadata Prediction  & 78.33 & 86.67 & 93.33 & 63.33 & 93.33 & 90.00 & \textbf{95.00} \\
Emotion Recognition  & 28.33 & 27.50 & 32.50 & 21.67 & 25.83 & 21.67 & \textbf{34.17}  \\
Composer Recognition & 30.21 & \textbf{36.46} & 32.29 & 20.83 & 27.08 & 27.08 & 30.21  \\
Genre Recognition    & 22.73 & \textbf{34.85} & 31.82 & 27.27 & 24.24 & 25.76 & 30.30 \\
\textbf{Sequence-level Average}   & 37.25 & 41.76 & 43.65 & 32.29 & 42.10 & 38.24 & \textbf{44.60}  \\
\midrule
\textbf{Overall Average}     & 40.17 & 52.01 & 51.25 & 44.14& 53.51 & 48.98 & \textbf{55.02} \\
\bottomrule
\end{tabularx}
\end{table*}

\vspace{-2mm}
\section{Experiments}
\vspace{-2mm}

This section reports a wide variety of experiments 
 conducted for evaluating 
 the comprehensive music understanding abilities of LLMs
 on the proposed ABC-Eval benchmark.

\vspace{-2mm}
\subsection{Experimental conditions}
\vspace{-1mm}

To assess the state-of-the-art performance in text-based symbolic music tasks, we use seven LLMs including three major families, e.g., DeepSeek-reasoner, DeepSeek-chat~\cite{DeepSeekr1,deepseek}, Gemini-2.5-flash, Gemini-2.5-pro~\cite{gemini}, GPT-5, GPT-5-mini and GPT-5-nano~\cite{GPT5}. We accessed these models via their official API services, primarily because these commercial LLMs represent the cutting-edge, allowing us to evaluate the most advanced capabilities of existing LLMs.

To minimize the interference of random factors on test results, we set the $temperature$ parameter to 0 and the $top\_p$ sampling parameter to 1, effectively disabling the models' random sampling functionality. During test sample input, we employed identical prompt templates for all models. For fair comparison, we did not utilize roleplay or chain-of-thought (CoT) prompting techniques. The specific prompt templates we used are shown in Table~\ref{tab:prompt}. Our input consists of the ABC score, task instruction, options or templates defining the output format, and simple natural language instructions.

\vspace{-2mm}
\subsection{Experimental results}
\vspace{-1mm}

The evaluation results of each baseline are listed in Table~\ref{tab:multi_task_eval}. 
Most foundation models exceeded 90\% accuracy on basic syntax understanding tasks, indicating their potential in symbolic music tasks. However, certain failure patterns (e.g., 30\% accuracy achieved by DeepSeek-chat model) reveal that performances on even the simplest tasks are highly sensitive to prompt variations, indicating a significant lack of robustness. 

Segment-level tasks proved more challenging, as all foundation models failed to tackle them effectively. In contrast, ultra-large or mixture-of-experts (MoE) models consistently demonstrated superior performance over their lighter-weight counterparts. This clear performance gap validates that the benchmark can effectively discriminate symbolic music understanding capabilities between models.

The sequence-level evaluation results reveal that complex musical reasoning tasks remain a great challenge for text-based LLMs, as evidenced by performances close to random guessing (25\%) on emotion/composer/genre recognition tasks. However, their superior performance on the Metadata Prediction task confirms that these models are fully capable of basic music reasoning, suggesting their potential in deep music reasoning tasks. 

Among all evaluated foundation models, GPT-5 achieved the best performance across segment-level, sequence-level, and overall tasks, demonstrating state-of-the-art capabilities in symbolic music understanding and instruction-following among general-purpose LLMs. Ultra-large or MoE models (e.g., DeepSeek-reasoner and Gemini-2.5-pro) also delivered promising overall performance.
Notably, performance comparisons within individual model families (e.g., GPT-5 versus GPT-5-nano) revealed that LLMs’ symbolic music understanding capabilities are strongly influenced by their parameter size and reasoning capacity. This observation provides valuable guidance for developing music-specific LLMs.

\vspace{-2mm}
\section{Conclusion}
\vspace{-2mm}

% In this work, we presented ABC-Eval, a novel and comprehensive benchmark for evaluating text-based LLMs on symbolic music understanding and instruction-following. The ABC-Eval benchmark comprises 1,086 test samples across ten subtasks and stands as the first benchmark dedicated to text-based symbolic music. We evaluated seven state-of-the-art LLMs on this benchmark, the experimental results underscore a substantial gap between the theoretical potential of LLMs and their practical capabilities in this domain. The variation in performance with model parameter size also highlights the importance of adopting foundation models with larger parameter scales and stronger reasoning capabilities. Future work will focus on designing finer-grained instruction-following tasks to improve the benchmark, and thus develop a better LLM framework for symbolic music application.

This study introduces ABC-Eval, 
 a pioneering and comprehensive open-source benchmark 
 designed to rigorously assess 
 the symbolic music understanding 
 and instruction-following capabilities of text-based LLMs. 
% It is the first benchmark specifically 
%  dedicated to evaluating LLMs on text-encoded symbolic music, 
%  leveraging the lightweight and human-readable ABC notation 
%  as its input format. 
Comprising 1,086 diverse test samples across 10 distinct sub-tasks, 
 it spans a broad spectrum of musical comprehension, 
 from fundamental syntax parsing 
 and segment-level understanding to sequence-level reasoning.
Our extensive evaluation
 revealed notable limitations 
 in the high-level understanding capabilities of LLMs. 
% While models showed proficiency in basic syntax understanding, 
%  their performance was sensitive to prompt variations, 
%  indicating a lack of inherent robustness. 
% Crucially, segment-level tasks proved challenging for all models, and complex 
% sequence-level reasoning tasks, such as emotion, composer, and genre recognition, often yielded performances barely above random guessing, highlighting a substantial gap between LLMs' theoretical potential and practical capabilities in this specialized domain. 

ABC-Eval could serve as a vital tool for advancing music research 
 in text-based symbolic music understanding by LLMs. 
Future work will focus on refining the benchmark 
 with even finer-grained instruction-following tasks 
 and leveraging these insights to develop 
 more robust and semantically sophisticated LLM frameworks 
 specifically tailored for symbolic music applications

{
\small
\bibliographystyle{IEEEbib_abrv}
\bibliography{strings,AANabrv,IEEEabrv,refs}
}
\end{document}